%% file: gabor_arxiv.tex
\documentclass{article}
\usepackage[preprint,nonatbib]{neurips_2023}
\usepackage[utf8]{inputenc} 
\usepackage[T1]{fontenc}    
\usepackage{hyperref}       
\usepackage{url}            
\usepackage{amsfonts}       
\usepackage{nicefrac}       
\usepackage{microtype}      
\usepackage{xcolor}         
\usepackage{booktabs}       
\usepackage{multicol}
\usepackage{multirow}
\usepackage{wrapfig}

\usepackage{amsmath,amsthm,mathtools,amsfonts,amssymb}
\usepackage{graphicx}
\usepackage{algorithm, algorithmic}
\usepackage{subfigure,subfloat}
\usepackage{soul}

\title{GaborPINN: Efficient physics informed neural networks using multiplicative filtered networks}
\author{Xinquan Huang, Tariq Alkhalifah\\
King Abdullah University of Science and Technology\\
\{xinquan.huang, tariq.alkhalifah\}@kaust.edu.sa}

\begin{document}

\maketitle

\begin{abstract}
The computation of the seismic wavefield by solving the Helmholtz equation is crucial to many practical applications, e.g., full waveform inversion. 
Physics-informed neural networks (PINNs) provide functional wavefield solutions represented by neural networks (NNs), but their convergence is slow. 
To address this problem, we propose a modified PINN using multiplicative filtered networks, which embeds some of the known characteristics of the wavefield in training, e.g., frequency, to achieve much faster convergence. 
Specifically, we use the Gabor basis function due to its proven ability to represent wavefields accurately and refer to the implementation as GaborPINN. 
Meanwhile, we incorporate prior information on the frequency of the wavefield into the design of the method to mitigate the influence of the discontinuity of the represented wavefield by GaborPINN. 
The proposed method achieves up to a two-magnitude increase in the speed of convergence as compared with conventional PINNs.
\end{abstract}
\input{introduction}
\input{method}

\input{experiments}

\input{discussion}
\input{conclusion}
\input{ack}

\bibliographystyle{IEEEtran}
\bibliography{samplebib}
\end{document}

%% file: introduction.tex
\section{Introduction}
Seismic wavefield simulation is a crucial and computationally intensive part of the many seismic imaging problems, e.g., reverse time migration and full waveform inversion (FWI).
An efficient simulation approach is quite critical to practical applications. 
Compared to the time domain,  frequency-domain modeling is often more efficient for applications like FWI \cite{Marfurt1984,GerhardPratt1998}. However, the solution requires the calculation of the inverse of the impedance matrix, which consumes a lot of memory, and this problem becomes more drastic as the model size increases, like in 3D. 
Furthermore, dense discretization is required for high-accuracy simulation when dealing with irregular geometry, e.g., topology and complex subsurface structures. 
With the recent developments in machine learning in science and engineering (so-called scientific machine learning), one type of approach, which embeds physical knowledge into the training, named physics-informed neural network (PINN), has provided the potential to solve this problem \cite{Raissi2019}. 
Specifically, we could use a neural network (NN) to represent the wavefield as a function of space (and time), train it to satisfy the governing equation, and the simulation happens on the query with input given by space coordinates and time, and the output given by the value of the wavefield.
This machine-learned function form of the wavefield \cite{Alkhalifah2021,waheed2022kronecker,song2023simulating} allows for easy handling of irregular geometry and can adapt to more complex wave equations corresponding to more complex media \cite{song_versatile_2021}.

However, the scalability of the approach is limited by the cost of the training for PINNs \cite{moseley_finite_2021}. 
Specifically, the PINNs training often provides solutions for one instance (e.g., one velocity model), and the convergence of each training may require thousands of epochs, making the total computational cost less competitive than numerical methods. 
So, {\it how can we make PINNs learn faster is an interesting, challenging, but inevitable topic.} 
As we know, there are three main components in PINNs: the neural network architecture, the training process, and the loss function \cite{cuomo_scientific_2022}. 
As for the training process, Huang and Alkhalifah 
\cite{huang_pinnup_2022} proposed a frequency upscaling and neuron splitting algorithm, resulting in a more stable and faster convergence, and 
Waheed {\it et al.} 
\cite{Waheed2021} proposed to use transfer learning to improve the computational efficiency by reducing the epochs needed for the convergence when applied to new velocity models.
As for the loss function design, 
Xiang {\it et al.} 
\cite{xiang_self-adaptive_2022} proposed a self-adaptive loss function through the adaptive weights for each loss term to adjust the collocation point samples in the domain to improve the accuracy of PINNs.
Huang and Alkhalifah 
\cite{huang_single_2022} proposed a single reference frequency loss function to improve the convergence of the multi-frequency wavefield representation. 
While for the neural network architecture design, almost all the backbone architecture used is the vanilla MLP with different activation functions. In this paper, we focus on the development of this aspect. 

To make the neural network fit the wavefield faster, prior knowledge should be embedded in the design of the PINN. 
For example, the Gabor function has been shown to effectively represent the seismic wavefield \cite{womack1994seismic}. 
Inspired by this fact, we include the Gabor function into the neural network by means of the Multiplicative filtered network (MFN) \cite{fathony2020multiplicative} to accelerate the convergence of PINNs. 
We propose a modified PINN with MFN and we refer to this network as GaborPINN. 
In this framework, we represent the wavefield by a linear combination of Gabor basis functions of the input coordinates (GaborNet), in which the scale factor is determined by the frequency of the wavefield. 
The prior information on this combination would be beneficial for the convergence of the fitting as the seismic wavefield could naturally be represented as a linear combination of basis functions \cite{kennett1994representations}, e.g., Gabor basis.  
Although Fathony {\it et al.}
\cite{fathony2020multiplicative} mentioned that this type of NNs retains some drawbacks such as the lack of smoothness in the represented function and its gradients; we found that with the proper scale selection for the Gabor function, this problem can be avoided.
We demonstrate the advantages of the method on a simple layered model extracted from the
Marmousi model and also discuss the scale factor selection. 
Further experiments on higher-frequency wavefields show that the proposed method results in faster convergence and provides higher accuracy where the vanilla PINN fails.

%% file: method.tex
\section{Methodology}
The framework of PINNs aims to train an NN function by using the governing equations of the physical system as a loss function. 
Here, we take the Helmholtz equation for a scattered wavefield \cite{Alkhalifah2021} as an example,
\begin{equation}
\frac{\omega^2}{\mathbf{v}^2} \delta \mathbf{U}+\nabla^2 \delta \mathbf{U}+\omega^2 (\frac{1}{\mathbf{v}^2}-\frac{1}{\mathbf{v}^2_0})\mathbf{U}_0=0,
    \label{equ:scq}
\end{equation}
where $\mathbf{U}_0$ is the background wavefield analytically calculated for a constant velocity $\mathbf{v}_0$ \cite{Richards1980}:
\begin{equation}
\mathbf{U}_0(x, z)=\frac{i}{4} \boldsymbol{H}_0^{(2)}\left(\omega \sqrt{\frac{\left\{\left(x-s_x\right)^2+\left(z-s_z\right)^2\right\}}{\mathbf{v}^2_0}}\right),
\end{equation}
where $\boldsymbol{H}_0^{(2)}$ is the zero-order Hankel function of the second kind, $\delta \mathbf{U}$ is the scattered wavefield and $\delta \mathbf{U}=\mathbf{U}-\mathbf{U}_0$, $\mathbf{v}$ is the velocity, and $\omega$ is the angular frequency.
Unlike the full wavefield, the scattered wavefield helps us avoid the point source singularity \cite{Alkhalifah2021}.
We define an NN function $\Phi(\mathbf{x})$ to map from the input coordinates to the output scattered wavefield value for the input location, where $\mathbf{x}=\{x,z,s_x\}$ (in 2D case, and sources are considered on the surface) represents the spatial coordinates and source location. 
To train an NN to satisfy the governing equation~\ref{equ:scq}, we evaluate the PDE residuals, given the input coordinates and the output wavefield of the NN, as a loss function to train the NN. 
Thus, the loss function is defined as
\begin{equation}
\mathcal{L}=\frac{1}{N} \sum_{i=1}^N\left|\frac{\omega^2}{(\mathbf{v}^i)^2} \Phi\left(\mathbf{x}^i\right)+\nabla^2 \Phi\left(\mathbf{x}^i\right)+ 
(\frac{\omega^2}{(\mathbf{v}^i)^2}-\frac{\omega^2}{(\mathbf{v}_0^i)^2})U_0^i\right|_2^2,
    \label{equ:loss}
\end{equation}
where $U_0^i$ are samples of $\mathbf{U}_0$ at point $x^i$. With this loss function, the NN can be trained as an alternative to the seismic wavefield simulation.

However, using the loss function (Equation \ref{equ:loss}), without additional constraints, may allow the network to converge to trivial solutions, e.g., a solution proportional to the negative background wavefield.
Inserting $\Phi=-\mathbf{U}_0$ into equation~\ref{equ:loss}, yields 
\begin{equation}
    \begin{aligned}
\mathcal{L}_{t}&=\frac{1}{N} \sum_{i=1}^N\left|-\frac{\omega^2}{(\mathbf{v}^i)^2} U_0^i -\nabla^2 U_0^i+\omega^2 
(\frac{1}{(\mathbf{v}^i)^2}-\frac{1}{(\mathbf{v}_0^i)^2})U_0^i\right|_2^2\\
&=
\frac{1}{N} \sum_{i=1}^N\left|-\nabla^2 U_0^i+\omega^2 
(-\frac{1}{(\mathbf{v}_0^i)^2})U_0^i\right|_2^2.
    \label{equ:loss_u0}
    \end{aligned}
\end{equation}
Since the background wavefield $\mathbf{U}_0$ satisfies the Helmoholtz equation:
\begin{equation}
\left(\frac{\omega^2}{\mathbf{v}_0^2}+\nabla^2\right) \mathbf{U}_0(\mathbf{x})=\mathbf{s},
    \label{equ:helm}
\end{equation}
where $\mathbf{s}$ is the point source, $\Phi=-U_0$ is a trivial solution.. 
Thus, the value of the loss $\mathcal{L}_{t}$,
for most collocation points (samples that are not close to the source) in the domain, is equal to zero for the trivial solution.
As the NN is trained to minimize the loss function for all collocation points, training a network without proper initialization or other strategies (like PINNup \cite{huang_pinnup_2022}) would push the NN to the trivial solution.
In this paper, to make the training of PINN stable and avoid the trivial solution, we propose a soft constraint (penalty term) to push the NN to the right path away from the trivial solution.
Considering that the value of the scattered wavefield near the source location is close to zero, while the value of the background wavefield near the source is often large, we add a penalty term that regularizes the predicted wavefield near the source (the region covering an area of almost one wavelength away from the source location). This penalty is given by
\begin{equation}
    \mathcal{L}_{reg}=\frac{1}{N_{reg}} \sum_{i=1}^{N_{reg}}\left| \Phi(\mathbf{x}^i)\right|_2^2.
    \label{equ:loss_reg}
\end{equation}

\begin{figure*}
    \centering
    \includegraphics[width=\textwidth]{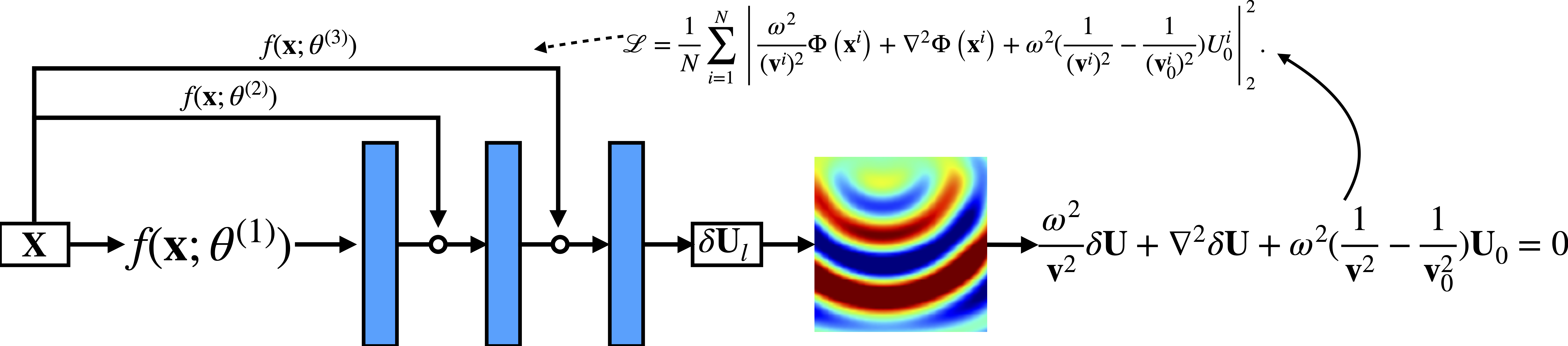}
    \caption{The GaborPINN framework for seismic wavefield simulation.}
    \label{fig:diagram}
\end{figure*}

In the vanilla PINN, the backbone NN is a Multilayer perception (MLP), given by the following form
\begin{equation}
\mathbf{h}^{(i+1)} =\sigma\left(\mathbf{W}^{(i)} \mathbf{h}^{(i)}+\mathbf{b}^{(i)}\right), i=1, \ldots, L-1,
\end{equation}
where $\mathbf{h}^{(i)}$ is the hidden layer output of layer $i$, $\sigma$ is a nonlinear activation function (we use the sine function here), $\mathbf{W}$ is the weight matrix, and $\mathbf{b}$ is the bias vector.
In our method, we use the multiplicative filter network \cite{fathony2020multiplicative}, which uses a different recursion that never results in the composition of nonlinear functions.
The hidden layer, indexed by $i$, is defined as
\begin{equation}
\mathbf{h}^{(i+1)} =\left(\mathbf{W}^{(i)} \mathbf{h}^{(i)}+\mathbf{b}^{(i)}\right) \circ f\left(\mathbf{x} ; \theta^{(i+1)}\right), i=1, \ldots, k-1 
\end{equation}
where $f$ is the filter function parameterized by $\theta$ directly applied to the input $\mathbf{x}$, and $\circ$ is an elementwise multiplication. 
Specially, we use the Gabor function here (GaborNet) because it often well represents the wavefield as a basis function, and is given by 
\begin{equation}
    f_j\left(\mathbf{x} ; \theta^{(i)}\right)=\exp \left(-\frac{\gamma_j^{(i)}}{2}\left\|\mathbf{x}-\boldsymbol{\mu}_j^{(i)}\right\|_2^2\right) \sin \left(\boldsymbol{\omega}_j^{(i)} \mathbf{x}+\phi_j^{(i)}\right)
    \label{equ:gabor}
\end{equation}
where $\theta^{(i)}$ is a group of parameters that control the shape of the Gabor kernel, including $\gamma_j^{(i)},\mu_j^{(i)},\omega_j^{(i)},\phi_j^{(i)}$, and $j$ is the index of the neurons in layer $i$.
Thus, we refer to this implementation combined with PINN as GaborPINN, in which a diagram of it is shown in Figure~\ref{fig:diagram}. 
For the first layer, $\mathbf{h}^{(1)} =f\left(\mathbf{x} ; \theta^{(1)}\right)$.
\cite{fathony2020multiplicative} mentioned that the MFN is generally rougher (less smooth) than the conventional MLP in representation and gradient calculation. 
This is because the wavenumber for the sine plane wave, $\boldsymbol{\omega}^{(i)}_j$, in equation \ref{equ:gabor} end up being very large.
So, we propose here to connect the hyperparameter $\boldsymbol{\omega}_j^{(i)}$ to the frequency of the wavefield. 
Specifically, we initialize $\boldsymbol{\omega}_j^{(i)}$ using
\begin{equation}
    \boldsymbol{\omega}_j^{(i)} = \{\omega_{scale} * \sqrt{\gamma_j^{(i)}},\omega_{scale} * \sqrt{\gamma_j^{(i)}},\omega_{scale} * \sqrt{\gamma_j^{(i)}}\}
    \label{w_init}
\end{equation}
where $\omega_{scale}$ is the scale factor to control the amplitude of the initialization of $\omega_j^{(i)}$.
For different input coordinates, their initializations are the same but will be updated to different values after training.
We discuss the selection of this scale factor later.

%% file: experiments.tex
\section{Examples}
\begin{figure}[!htb]
    \centering
    \includegraphics[width=1.0\columnwidth]{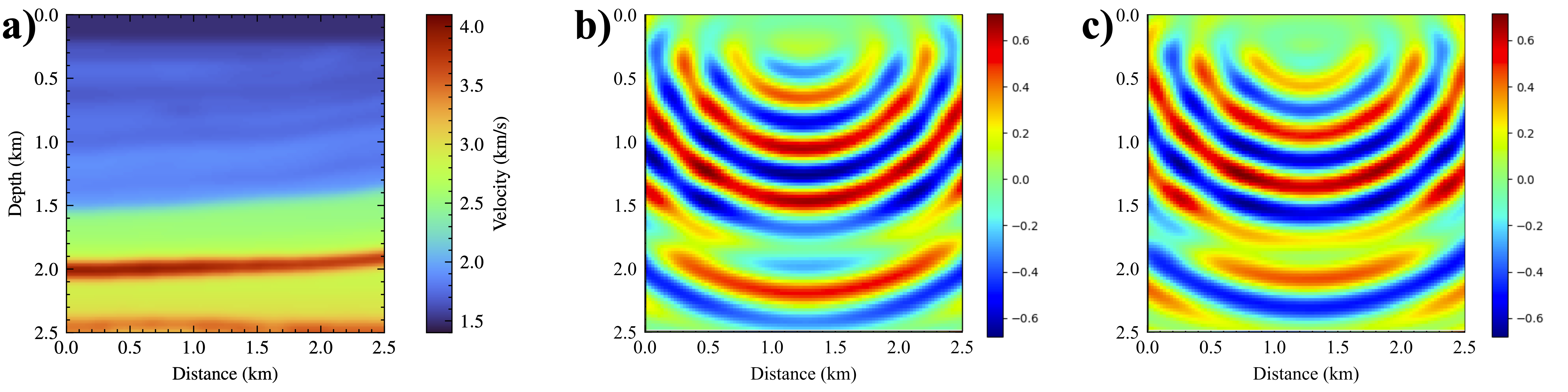}
    \vspace{-6pt}
    \caption{True velocity (a); the real (b) and imaginary (c) parts of the 4 Hz scattered wavefield calculated numerically.}
    \label{fig:numer}
\end{figure}

In this section, we will test three versions of PINNs to evaluate the convergence and accuracy improvements of GaborPINN. 
The tests are based on a simple 2.5$\times$2.5 $km^2$ layered model extracted from the Marmousi model (Figure~\ref{fig:numer}). 
We use 40000 random samples from this region for training, and each sample is given by the spatial coordinates $x$, $z$ for the wavefield, $x_s$ for the location of the source near the surface, velocity $v$, and a constant background velocity of 1.5 $km/s$.
Actually, the depth of the sources is fixed at 0.025 $km$.
We train the network for a frequency-domain wavefield of 4 Hz using an Adam optimizer for 50000 epochs. To evaluate the results, we solve the Helmholtz equation using the finite-difference method for a frequency of 4 Hz with using a fine grid spacing for accuracy to act as a reference. 
In the following experiments, we show the results of MLP and GaborPINN comprised of 3 hidden layers with 256 neurons in each layer, and a larger MLP comprised of 3 hidden layers with 512 neurons in each layer.
A comparison with respect to network size and computational complexity is shown in Table~\ref{tab:c}.
\begin{table*}[ht]
    \centering
    \caption{The capacity and complexity comparison between MLP and GaborPINN}
    \setlength{\tabcolsep}{0.025\columnwidth}
    \begin{tabular}{c|c|c|c}
         \toprule
         NN type & Trainable parameters & \# & Computational Complexity \\
         \midrule
        MLP-256 & \{$\mathbf{W}^{(i)}, \mathbf{b}^{(i)}$\} & 133.12 k & 133.12 KMac \\
        GaborPINN & \{$\mathbf{W}^{(i)}, \mathbf{b}^{(i)}, \gamma_j^{(i)},\boldsymbol{\mu}_j^{(i)},\boldsymbol{\omega}_j^{(i)},\phi_j^{(i)} $\} & 206.08 k & 201.99 KMac\\
        MLP-512 & \{$\mathbf{W}^{(i)}, \mathbf{b}^{(i)}$\} & 528.39 k & 528.39 KMac\\
        \bottomrule
    \end{tabular}
    \label{tab:c}
\end{table*}

\begin{figure}
    \centering
    \includegraphics[width=0.85\columnwidth]{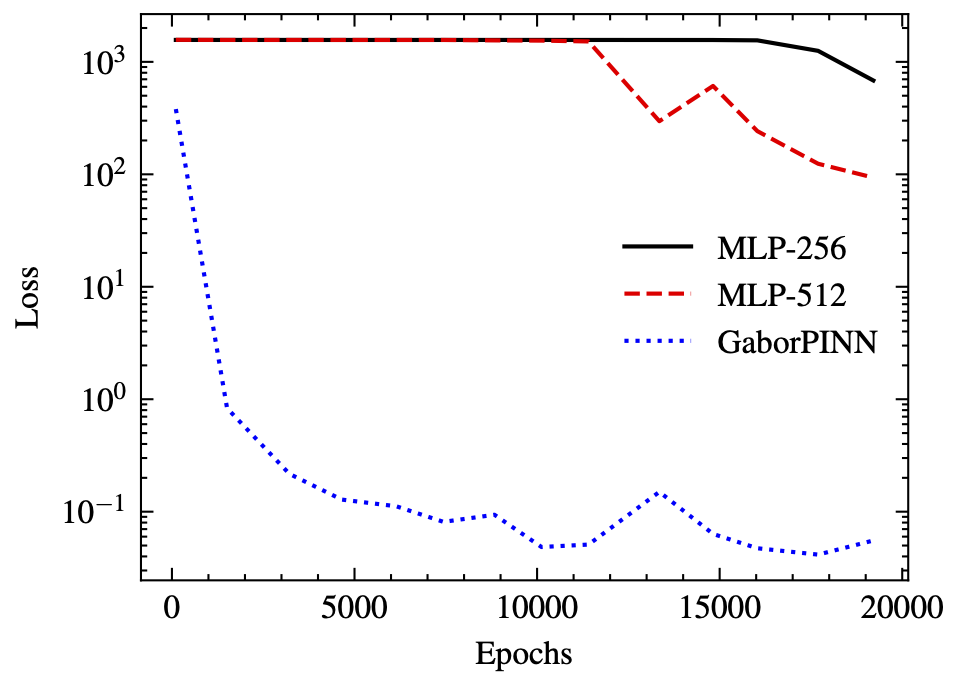}
    \caption{The Loss curves for three versions of PINNs.}
    \label{fig:4hz_loss}
\end{figure}

Figure~\ref{fig:4hz_loss} shows the loss curves for these three different NNs, in which the NNs are trained by the physical loss. 
GaborPINN has the best convergence by far compared to the other networks.
The loss function for the vanilla PINN stagnates for thousands of epochs and decreases slowly as it has no prior information related to the seismic wavefield. On the other hand, we could see that with GaborPINN, which incorporates the prior information of the wavefield into the network design, as well as inject the input coordinates at every neuron, the convergence is quite good and we could reduce the needed training epochs by two orders of magnitude. 

\begin{figure}
    \centering
    \includegraphics[width=\columnwidth]{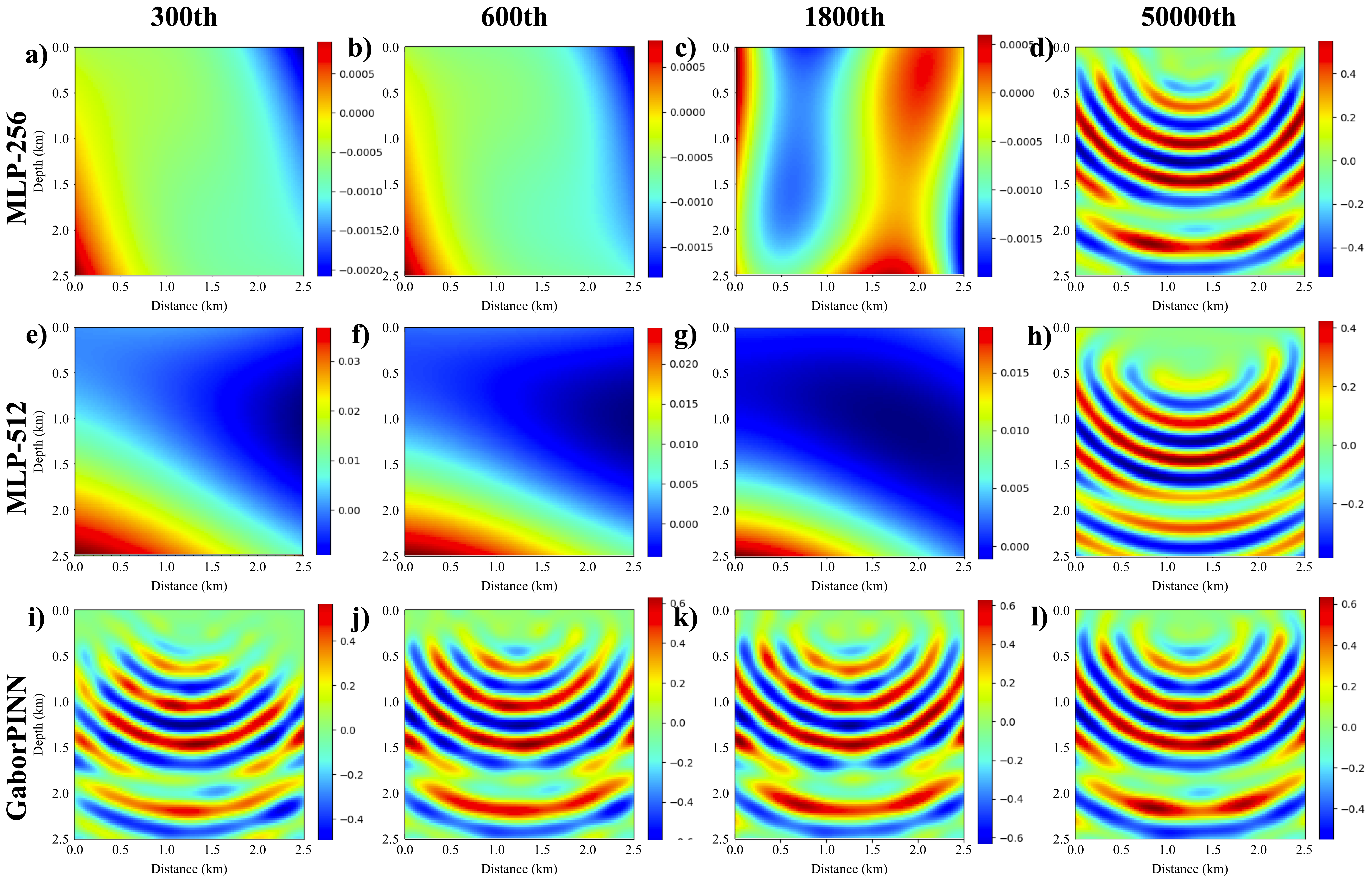}
    \caption{The real-part predictions of the 4Hz wavefield due to a source near the surface at 1.25 km at various epochs (b-m) for three versions of PINNs. (b-e) are the results for the MLP comprised of 3 hidden layers with 256 neurons in each layer, (f-i) are the MLP comprised of 3 hidden layers with 512 neurons in each layer, and (j-m) are the results of GaborPINN.}
    \label{fig:result_pde}
\end{figure}

To better understand how many training epochs are needed for GaborPINN to provide a solution, we visualize the results at the 300th, 600th, 1800th, and 50000th epochs (Figure~\ref{fig:result_pde}). 
The GaborPINN could reconstruct the main parts of the wavefield within hundreds of epochs, and the details of the predictions are refined as the training progresses.
However, as for the vanilla PINN, the NN learns nothing in the first 1800 epochs, which is a common problem for wavefield solutions using PINNs.
Even when increasing the width of the NN in the vanilla PINN, the slow convergence persists.   

\begin{figure*}
    \centering
    \includegraphics[width=0.9\textwidth]{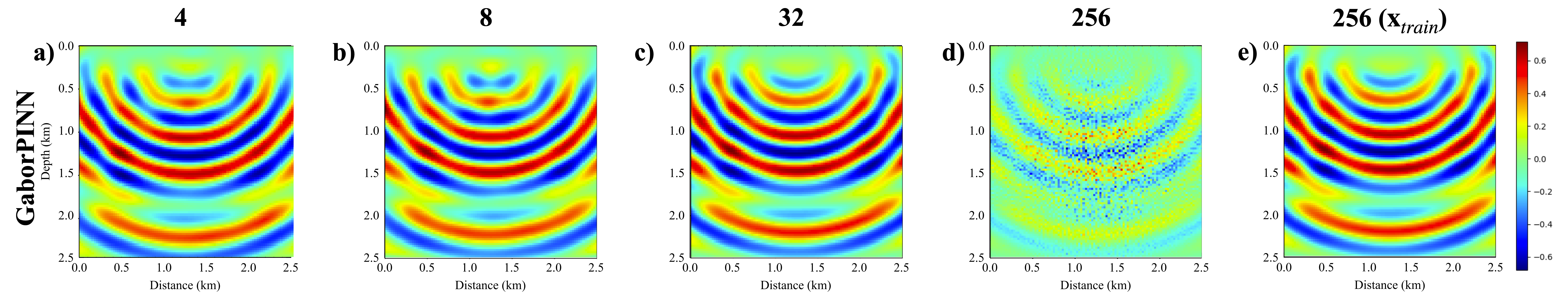}
    \vspace{-6pt}
    \caption{The prediction results (a-d) of GaborPINN with different initial frequency scales applied to testing points {$\mathbf{x}_{test}$}, which are perturbed compared to the training points {$\mathbf{x}_{train}$}, and that of GaborNet whose scale is 256 (e) evaluated at the training points {$\mathbf{x}_{train}$}.}
    \label{fig:result_mse}
\end{figure*}

The scale used here for the initialization of the $\boldsymbol{\omega}_j^{(i)}$ is 32, which differs from what was used in the original GaborNet. We found that a smaller value like 16 would introduce smoother, less accurate results, while a larger value would result in a failure of GaborPINN. 
So we test the GaborPINN for different values of initial spatial frequency (wavenumber) $\omega_j^{(i)}$ in the network under a supervised training setting on samples from a regular grid, ${\mathbf{x}_{train}}$ (Figure~\ref{fig:result_mse}), and use a grid that is slightly shifted from the training samples to show the wavefields. We found that a decrease in the value of $\omega_{scale}$ will make the representation smoother. However, for a value of 256, 
the predicted output wavefield is generally inaccurate. 
In other words, its smoothness or continuity is destroyed due to the network's focus on a high-frequency representation. 

\begin{figure}[!htb]
    \centering
    \includegraphics[width=0.8\columnwidth]{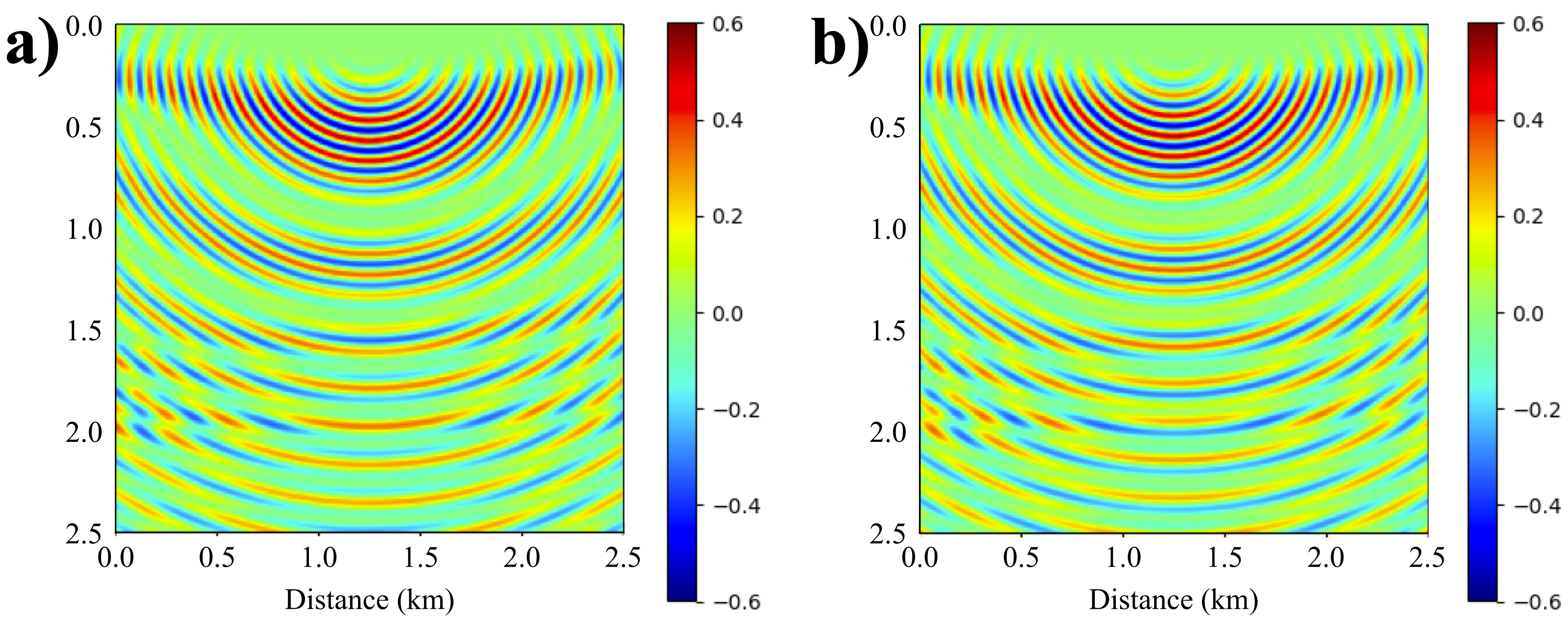}
    \vspace{-6pt}
    \caption{The real (b) and imaginary (c) parts of the 16 Hz wavefield calculated numerically.}
    \label{fig:numer_16hz}
\end{figure}

\begin{figure}
    \centering
    \includegraphics[width=0.85\columnwidth]{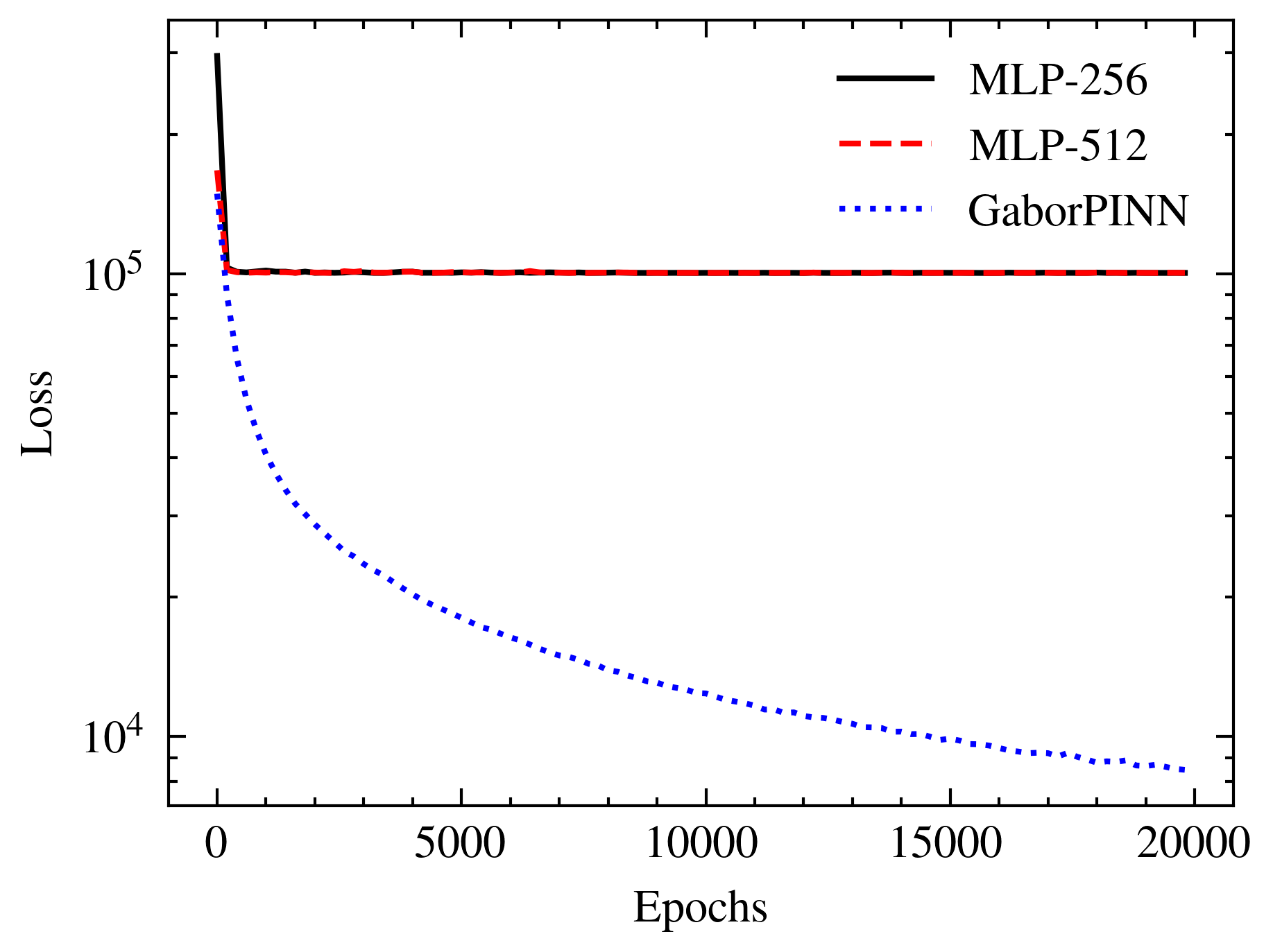}
    \caption{The Loss curves for three versions of PINNs.}
    \label{fig:16hz_loss}
\end{figure}

\begin{figure}
    \centering
    \includegraphics[width=\columnwidth]{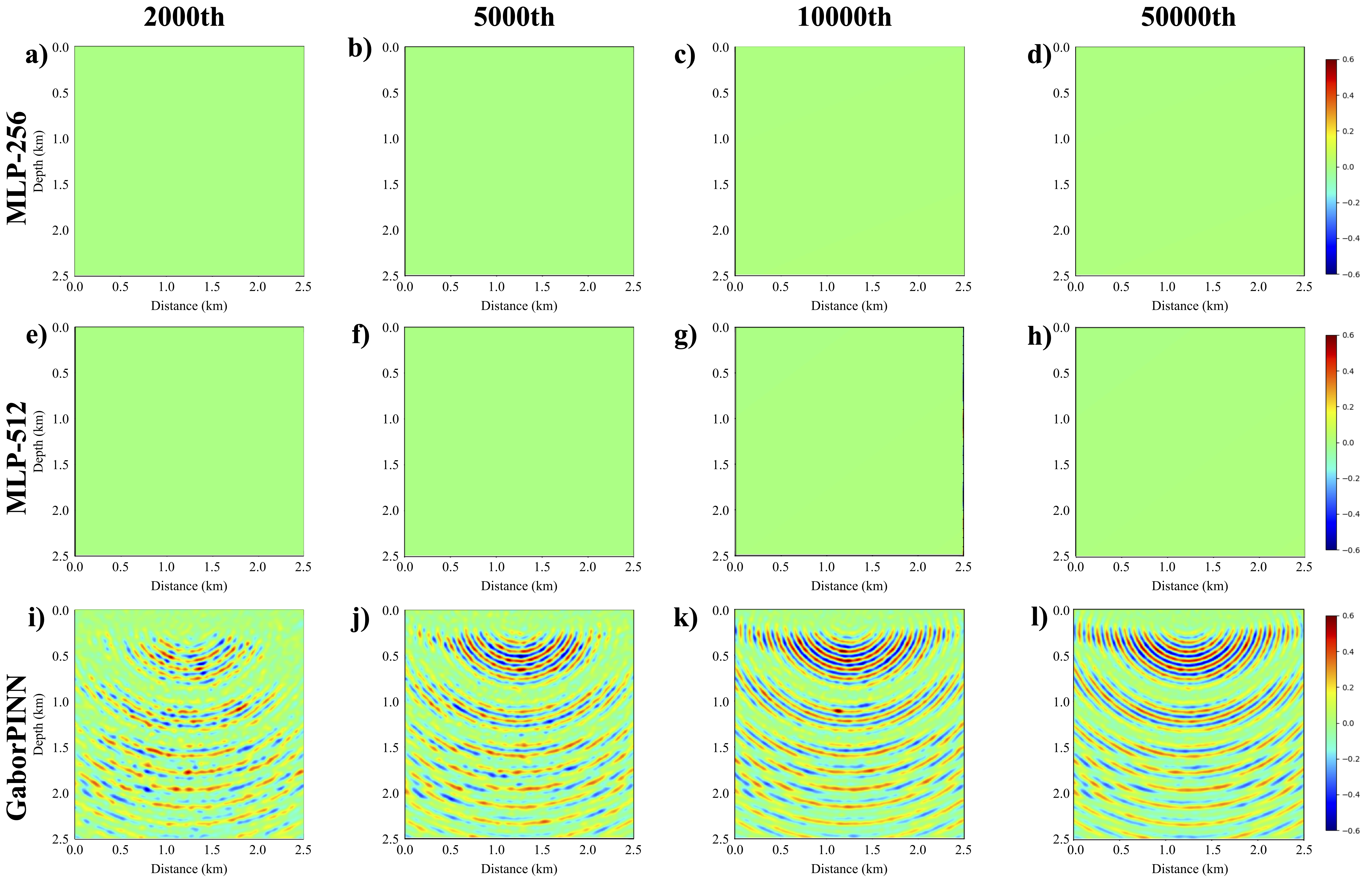}
    \caption{The predicted real part of the 16Hz wavefield due to a source near the surface at 1.25 km at various epochs (a-l) for three versions of PINN. (a-d) are the results for the MLP comprised of 3 hidden layers with 256 neurons in each layer, (e-h) are the MLP comprised of 3 hidden layers with 512 neurons in each layer, and (i-l) are the results of GaborPINN.}
    \label{fig:result_pde_16hz}
\end{figure}

Then we test the performance of GaborPINN in learning a 16 Hz seismic wavefield.
We still use the true velocity shown in Figure~\ref{fig:numer}a to generate the reference results (Figrue~\ref{fig:numer_16hz}) using finite difference methods.
The generation of the training samples is the same as the above experiment, but we use 160000 random samples from this region instead. 
We train the GaborPINN, the MLP with 3 hidden layers of 256 neurons in each layer, and the MLP with 3 hidden layers of 512 neurons in each layer, using an Adam optimizer for 50000 epochs. 
We note that for the GaborPINN, the scale factor used for the initialization of the $\omega_j^{(i)}$ is 128 to fit the high-frequency nature of the wavefield.
The loss curve is shown in Figure~\ref{fig:16hz_loss}.
The GaborPINN still performs well for a 16 Hz wavefield, while the vanilla PINN did not converge.
The predictions of the NN at different training epochs are shown in Figure~\ref{fig:result_pde_16hz}.
The GaborPINN learns this high-frequency wavefield with limited training epochs, and the details are refined later.
The vanilla PINN did not converge within 50000 epochs.
This example demonstrates that the GaborPINN provides fast convergence and a strong capability \cite{fathony2020multiplicative} to represent wavefields.
It is worth noting that the networks used for this high-frequency wavefield are the same in size as those used for the 4 Hz case, which probably hurt the conventional PINN. 
However, we the proper frequency scaling the GaborPINN managed to converge fast.

%% file: discussion.tex
\section{Discussion}
In this paper, we proposed a modified PINN for wavefield representation using GaborNet, which we refer to as GaborPINN. 
With the proper hyperparameter selection for GaborPINN, the convergence is fast and the prediction is good. This comes from the fact that 
the wavefield is inherently well represented by a composition of Gabor basis functions. 
Also, for this reason, the selection of an initial frequency scale is crucial. 
Unlike for image representation, which GaborNet is originally developed for, where they care about the details and the NN is trained in a supervised manner using the ground truth, PINNs with GaborNet require careful selection of the initial frequency scale. 
A large value will help GaborPINN fit the high-frequency details in the wavefield but will make the prediction rough, causing inaccurate second-order derivative calculations, which will harm the training of PINNs. 
Here, we choose the value for the scale based on the frequency of the wavefield, and as a result, we increase the initial frequency scale when we need wavefield solutions for higher frequencies.
For future research, {\it we will investigate a more intelligent or adaptive selection of the initial frequency scale}. 

%% file: conclusion.tex
\section{Conclusions}
We addressed the issue of the slow convergence of PINN-based wavefield solutions by incorporating Gabor basis functions. 
The Gabor function allowed us to incorporate the prior information on the frequency in the design of GaborPINN to accelerate the fitting of the NN to the Helmholtz equation. 
Unlike in image representation, we found that the proper choice of the initial frequency parameter in the Gabor function would highly affect the smoothness and continuity of the represented wavefield.
Therefore, we determine this scale value based on the frequency of the wavefield.
Through numerical tests, we show that the proposed approach converges much faster than vanilla PINNs. 
This method provides a stepping stone, from the perspective of NN design, to efficient wavefield representations for real problems using PINNs. 

%% file: ack.tex
\section*{Acknowledgements}
The authors thank KAUST 
for supporting this research, and Fu Wang for helpful discussions. We would also like to thank the SWAG group for the collaborative environment.